\documentclass[pre,twocolumn,superscriptaddress]{revtex4-1}
\usepackage{amssymb,amscd,amsmath}
\usepackage{graphicx}
\usepackage{color}
\usepackage{mathtools}
\def\tm{truemm}
\def\mb{\mbox}

\def\hsp#1{\hspace{#1 \tm}}

\def\hsa{\mbox{} \hsp{2} }
\def\hsb{\mbox{} \hsp{4} }
\def\beqa{\begin{eqnarray}}
\def\eeqa{\end{eqnarray}}
\def\a={&=&}
\def\b={\ = \ }
\def\nnn {\nonumber \\}
\def\pkcz{P^0_{s_k\to_{C}}}
\def\pkc{P_{s_k\to_{C}}}

\def\xc{x_{\!_C}}
\def\cth{c\sube{th}}

\def\rij{r_{\!_{ij}}}

\newcommand{\sube}[1]{_{\mb{\scriptsize #1}}}

\newcommand{\pa}[1]{\left( {#1} \right)}
\newcommand{\pas}[1]{\left[ {#1} \right]}

\begin{document}
\title[]{Long-range prisoner's dilemma game on a cycle}
\author{Jiwon Bahk}
\affiliation{Department of Physics and Astronomy, Sejong University, Seoul
05006, Korea}
\author{Seung Ki Baek}
\email[]{seungki@pknu.ac.kr}
\affiliation{Department of Physics, Pukyong National University, Busan 48513,
Korea}
\author{Hyeong-Chai Jeong}
\email[]{hcj@sejong.ac.kr}
\affiliation{Department of Physics and Astronomy, Sejong University, Seoul
05006, Korea}

\begin{abstract}
We investigate evolutionary dynamics of altruism with long-range
interaction on a cycle. The interaction between individuals
is described by a simplified version of the prisoner's dilemma (PD) game
in which the payoffs are parameterized by $c$, the cost of a cooperative
action.
In our model, the probabilities of the game interaction and competition
decay algebraically with $r_{AB}$, the distance between two players $A$ and $B$,
but with different exponents: That is, the probability to play the PD game is
proportional to $r_{AB}^{-\alpha}$. If player $A$ is chosen for death, on the
other hand, the probability for $B$ to occupy the empty site is
proportional to $r_{AB}^{-\beta}$.
In a limiting case of $\beta\to\infty$, where the competition for an empty site
occurs between its nearest neighbors only,
we analytically find the condition for the proliferation
of altruism in terms of $\cth$, a threshold of $c$ below which altruism
prevails. For finite $\beta$, we conjecture a formula for $\cth$ as a function
of $\alpha$ and $\beta$. We also propose a numerical method to locate $\cth$,
according to which we observe excellent agreement with the conjecture
even when the selection strength is of considerable magnitude.
\end{abstract}
\maketitle

\section{Introduction}

As Charles Darwin remarked in {\it On the Origin of Species},
``Natural selection will never produce in a being anything injurious
to itself, for natural selection acts solely by and for the good of each.''
The existence of altruism is seemingly at odds with this statement, as far as
altruists are defined as those who benefit others at the cost to themselves.
For this reason,
there have been extensive studies to explain the evolution of altruism.
The explanations can be categorized (arguably) into five rules, i.e., direct
reciprocity, indirect reciprocity, spatial reciprocity, kin selection,
and group
selection~\cite{nowak2006five,nowak2010evolution,*abbot2011inclusive,*marshall2011group}.
An essential common factor in these mechanisms is the
assortment of individuals carrying the cooperative
phenotype~\cite{fletcher2009simple,*park2012emergence,debarre2014social}.
Let us consider a population of defectors.
If there emerges a group within which the members cooperate
themselves,
its size will increase because the members earn
higher fitness than the population average. Furthermore,
if the assortment can be maintained while the size of the group grows,
the whole population will become cooperative.
On the other hand, if the assortment breaks, making
every individual interact with every other, cooperators
will disappear because they earn less than the population average by
construction.

Considering that the defectors always win in the mean-field regime,
one may expect that spatial reciprocity becomes weaker as the
dimensionality of the space $d$ increases. The logic goes as
follows: In a population of size $N$, the average chemical distance
between a pair of individuals behaves as $\sim N^{1/d}$ and thus
decreases dramatically as $d$ grows. Hence, it is reasonable to expect
a relatively homogeneous state in a high-dimensional structure.
In many statistical-physical models, the mean-field theory becomes
exact as $d$ exceeds a certain value~$d_c$\cite{landau1980statistical}.
Another way to look at the crossover between a low-dimensional spin
system and the mean-field limit is to consider long-range interactions
on a one-dimensional (1D) lattice 
structure~\cite{thouless1969long,*fisher1972critical,*luijten2000monte}. 
For example, when the interaction strength decays as $r^{-1-\sigma}$
between a pair of 1D Ising spins with distance $r$,
the spins show the mean-field-like critical behaviors for
$0 < \sigma \le 1/2$ as those with short range 
interactions in high dimensions $d>d_c=4$ do.   
For $1/2 \le \sigma \le 1$, 1D Ising spin system shows the
order-disorder transitions at finite temperatures but
its critical behaviors are different from mean-field transition.
For $\sigma>1$, the interactions become short-ranged effectively
and they behave as the 1D spins with finite-range interactions.  

The story becomes more complicated when it comes to
evolutionary games: Let us consider
the Prisoner's Dilemma (PD) game with the death-birth (DB)
reproduction process in the weak selection limit. It has a certain population
structure, and the game is played between players at one step away.
On regular structures, the condition for cooperation turns out to be
determined by the number of nearest neighbors regardless
of $d$~\cite{ohtsuki2006simple} (see also
Ref.~\onlinecite{allen2017evolutionary} for more general results).
Furthermore, spatial reciprocity breaks down for any population structures if
we employ the birth-death (BD) process instead.
What is the difference between the DB and BD processes? The point is that
we actually have another kind of interaction, that is,
competition for reproduction. For social behavior to evolve, these two should
have different length scales~\cite{debarre2014social}.
Even if the game interaction shares the same underlying structure with
competition for reproduction, the DB process naturally separates the scales
because the competition for reproduction effectively occurs between players at
\emph{two} steps away.

In this paper, we wish to investigate the roles of these two length scales
explicitly by considering the PD game on a cycle with long-range
interactions, whose dynamics is governed by the DB process.
The game is parameterized by the cost of cooperation, denoted by
$c$. We are interested in its threshold, $\cth$,
above which cooperators are outnumbered by defectors in the
infinite population with zero mutation limit. 
The length scale of the game interaction is determined by an exponent
$\alpha$, so that the game result between two individuals at distance $r$
has a weighting factor of $r^{-\alpha}$. On the other hand, the length scale of
competition for reproduction is determined by another exponent $\beta$,
so the probability for an individual to occupy an empty site at distance $r$
decays algebraically as $r^{-\beta}$.
Our first main result is to derive $\cth$ for $\beta \to \infty$, where
only the nearest neighbors compete for reproduction. Then, we
consider the case of $\alpha = \beta$, where an analytic result for $\cth$
is available in the weak-selection limit~\cite{allen2017evolutionary}.
Based on these two special cases, we conjecture a
general form of $\cth (\alpha, \beta)$. It agrees well with
numerical results.

This work is organized as follows: In the next section, we give a detailed
description of our model. In Sec.~\ref{sec:analysis}, we present $\cth$ as a
function of $\alpha$ in case of $\beta\to\infty$ and review the exact $\cth$
when $\alpha=\beta$ in the weak-selection limit. Based on these two results,
we conjecture $\cth$ for general $\alpha$ and $\beta$. Section~\ref{sec:numeric}
compares this conjecture with Monte Carlo data. The conjectured threshold nicely
matches with our numerical calculation, and this holds even if the selection
strength increases. We discuss this finding and conclude this work in
Sec.~\ref{sec:conclude}.

\section{Model}
The population structure is a cycle, i.e., a 1D lattice with the
periodic boundary condition. For the sake of convenience, the total
number of sites is assumed to be an add number in this section and we
set $N \equiv 2n+1$. Of course, it will be irrelevant
whether $N$ is even or odd if we deal with a sufficiently large system.
Every site~$i$ is occupied by a player, who is also denoted by
\mb{$i=-n$,} \mb{$-n+1$,} $\ldots, -1, 0, 1, \ldots, n-1, n$, and each player~$i$ can
choose an action $s_i$ between cooperation $C$ and defection $D$. The PD game is
formulated as the donation game, where a cooperator benefits the co-player by an
amount of $b > 0$ with paying cost $c \in (0,b)$. The payoff matrix $\pi$ is
thus defined as
\beqa
\begin{array}{cc}
  & \ \ \begin{array}{cc}
 \hsa C & \hsb\ D
      \end{array} \ \	\\
 \begin{array}{c}
      C      \\
      D
 \end{array}
 & \left( \begin{array}{cc}
      b-c & \hsa  -c   \\
      b   &  0
  \end{array} \right).
\end{array}
\label{eq.pi}
\eeqa
Without loss of generality, we will set $b=1$ throughout this work.
The expected payoff of player $i$ {\em per game} is given by the following
weighted sum, 
\beqa
\Pi_i
 \a= \sum_{j\not= i} \frac{\rij^{-\alpha}}{2\zeta_{n} (\alpha)} \pi_{s_i s_j}
 \b= \frac{1}{2\zeta_n (\alpha)}
 \sum_{j\not= i} \frac{ \pi_{s_i s_j}}{\rij^\alpha}
\label{e.Pi_i}
\eeqa
where $\zeta_{n}(\alpha) \equiv \sum_{k=1}^{n} k^{-\alpha}$ is the
incomplete Riemann zeta function and
$\rij^{-\alpha}/2\zeta_{n}(\alpha)$
is the probability that player~$i$ play a game with player $j$
where $\rij$ is the distance between them. On a cycle,
$\rij$ is given by the minimum of
$|i-j|$, $|i-j+N|$, and $|i-j-N|$. 
We consider the expected payoff per game (rather than the total
payoff, $\sum_{j\not= i} \rij^{-\alpha} \pi_{s_i s_j}$) 
to make the average payoff independent of the population size. 
This normalized payoff is also better to compare two systems with
different $\alpha$ fairly. The systems with smaller $\alpha$ would
show effectively more strong selections without the normalization.

\begin{figure}[t] 
\includegraphics[scale=0.5]{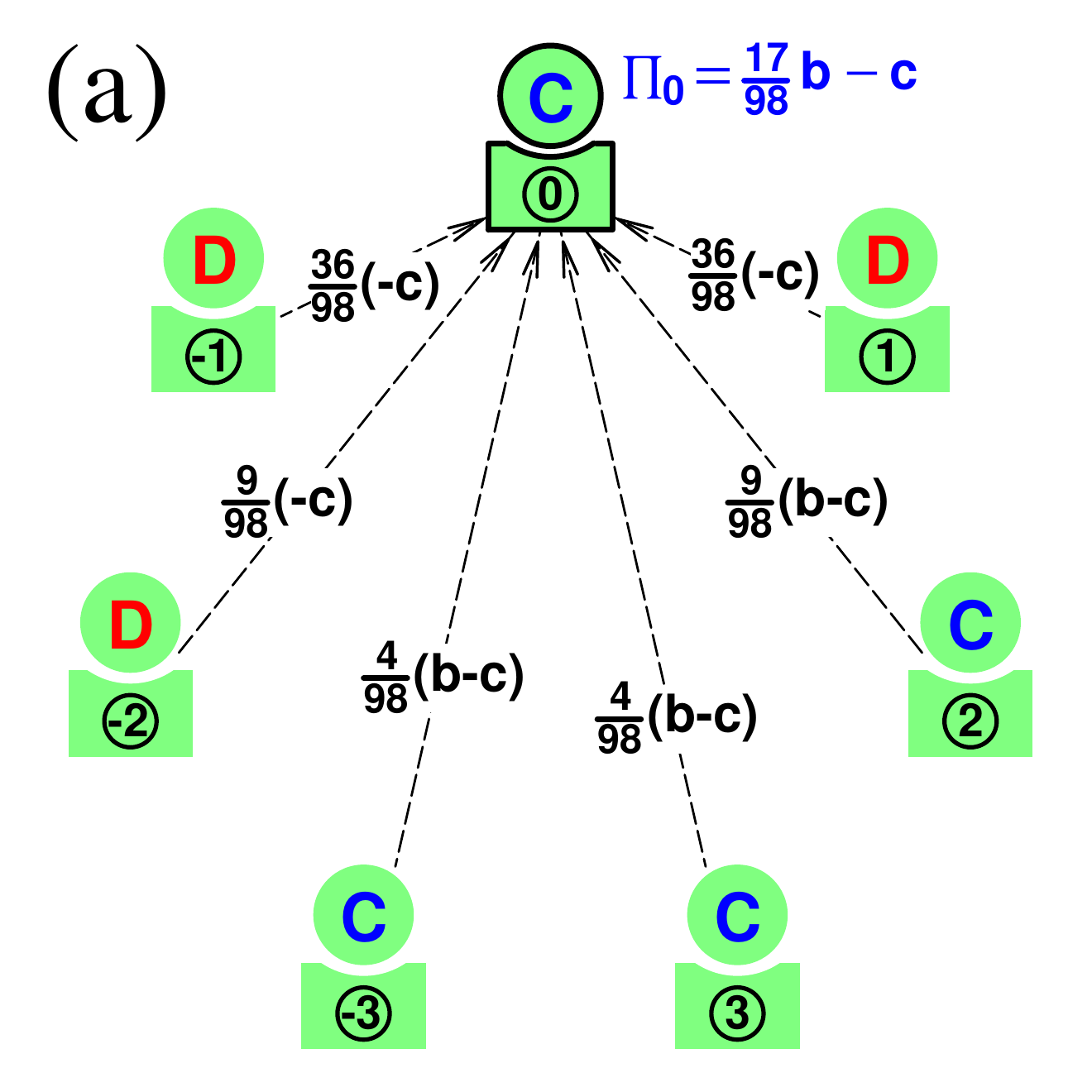}\\
\includegraphics[scale=0.5]{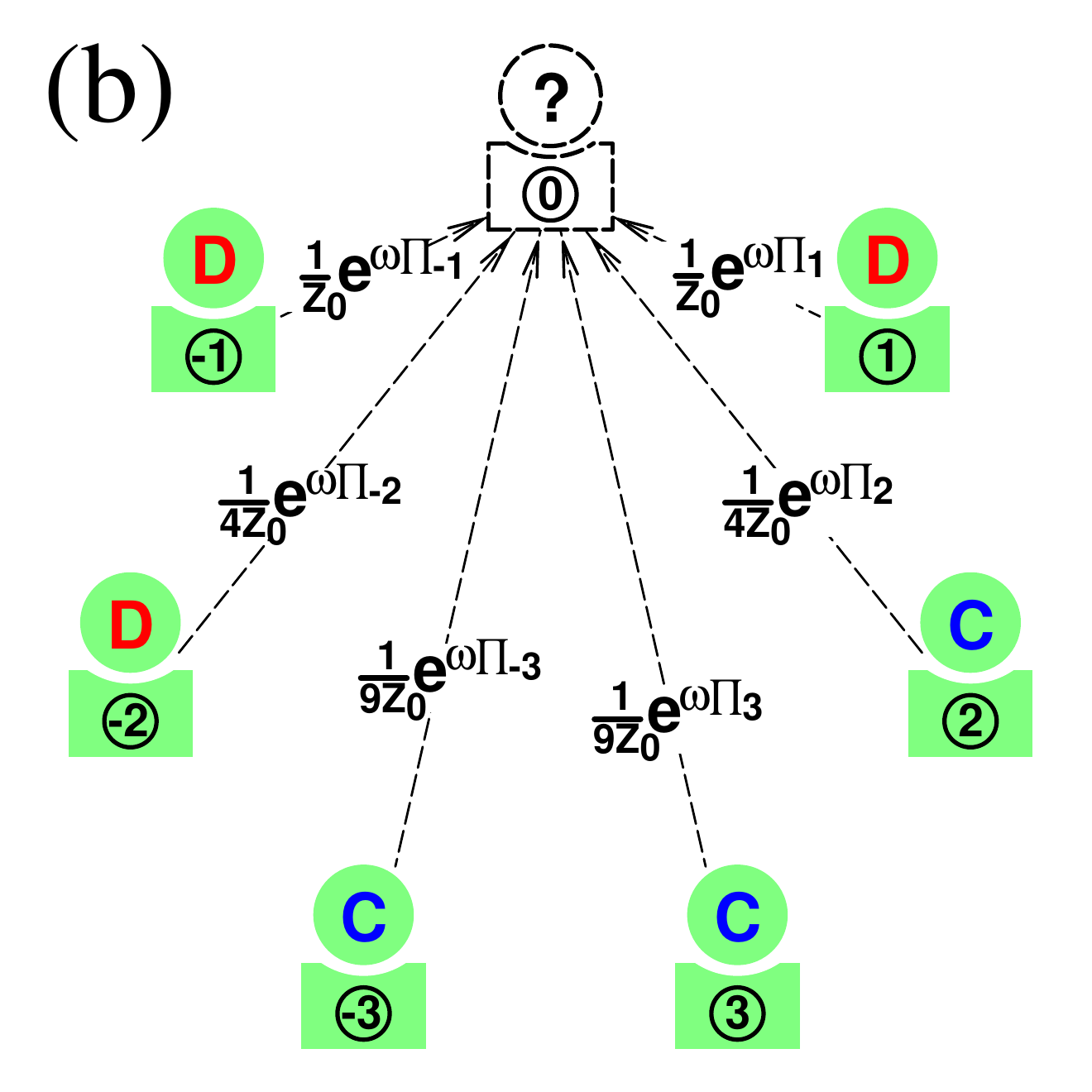}
\caption{(Color online) Schematic representation of an example
 of the model ($\alpha=\beta=2$, $n=3$).
(a) Payoff of the player~0.  
Player~$0$ plays the PD game of Eq.~~\eqref{eq.pi}
with player~$k\not=0$ producing
$\pi_{0k}=b-c$ for $k=-3$, 2, and 3 and 
$\pi_{0k}=-c$ for $k=-2, -1$ and $1$.
$\Pi_0$ is the weighted sum of
these values with the weighting factors, 
$\frac{|k|^{-2}}{2\zeta_3(2)}$.
Weighting factors are
$\frac{36}{98}$ for $n=\pm1$,
$\frac{9}{98}$ for $n=\pm2$, 
and $\frac{4}{98}$ for $n=\pm3$,
and Payoff of the player~0 is  
$\Pi_0= \frac{17}{98}b - c$.
Payoff $\Pi_k$ of player~$k\not=0$ 
can be obtained similarly;
$\Pi_{-3} = \frac{49}{98}b - c$,
$\Pi_{-2} = \frac{58}{98}b$,
$\Pi_{-1} = \frac{53}{98}b$,
$\Pi_{1} = \frac{85}{98}b$,
$\Pi_{2} = \frac{54}{98}b - c$,
$\Pi_{3} = \frac{49}{98}b - c$.
(b) Competition for reproduction. 
When a player, say, $i=0$, is chosen for death, the
other players $k\not=0$ compete to produce their offspring at this
empty site. Player $k$'s winning probability is proportional to
its fitness $f_k =e^{\omega \Pi_k}$ and inverse square distance
$|k|^{-2}$.  Note that $1/Z_0$ is to normalize the total sum of
probabilities. The winner's offspring inherits the winner's strategy
with probability $(1-\mu)$ or the opposite one with $\mu \ll 1$.
}
\label{f.model}
\end{figure} 

Fig.~\ref{f.model}(a) illustrates the way to calculate payoff
$\Pi_0$ of player~0 for $\alpha=2$ and $n=3$. 
Player~$0$ at the top plays the PD game with
every other in the population, $k=\pm3$, $\pm2$, and $k=\pm1$. 
From Eq.~\eqref{eq.pi}, we see that $\pi_{0k}$ is
$-c$ for $k=-2, -1$ and $1$ and $b-c$ for $k=-3$, 3, and 2.
$\Pi_0$ is the weighted sum of these values with the wighting factors,  
$\frac{|k|^{-2}}{2\zeta_3(2)}$ where
$2 \zeta_3(2) = 2\pas{1/1^2+ 1/2^2+1/3^2}= \frac{98}{36}$.  
Since $\frac{|k|^{-2}}{2\zeta_3(2)}$ is
$\frac{36}{98}$ for $n=\pm1$,
$\frac{9}{98}$ for $n=\pm2$, 
and $\frac{4}{98}$ for $n=\pm3$,
$\Pi_0$ is given by
$\pi_0= \frac{17}{98}b - c$.
We can calculate the payoffs of the other players similarly.
They are 
$\Pi_{-3} = \frac{49}{98}b - c$,
$\Pi_{-2} = \frac{58}{98}b$,
$\Pi_{-1} = \frac{53}{98}b$,
$\Pi_{1} = \frac{85}{98}b$,
$\Pi_{2} = \frac{54}{98}b - c$,
$\Pi_{3} = \frac{49}{98}b - c$.

Player $i$'s fitness is given by the exponential form of the payoff,
\beqa
f_i \a= e^{\omega \Pi_i},
\label{e.fi}
\eeqa
where the selection strength $\omega$ represents how strongly the
fitness depends on the payoff. Exponential fitness of Eq.~(\ref{e.fi})
becomes the usual linear fitness of $1+\omega \Pi_i$ in the
weak selection limit. It has an advantage over linear fitness in the
strong selection limit. We can interpret the exponential fitness
as fecundity directly even when the payoff is negative. Furthermore,
DB dynamics is invariant under the addition of a constant to all 
elements in the pay off matrix which certainly preserve the Nash
equilibrium.   

We employ a long-range version of the DB process as a model of
reproduction. The first step is to choose a death site $k$ randomly,
irrespective of fitness, and make it empty. The probability to choose
a particular site $k$ is therefore inversely proportional to $N$. 
The difference from the usual DB process for the nearest-neighbor
competition is that any player $i$ in the population can produce an
offspring at the empty site. For given $k$, the conditional
probability $p_{ik}$ that player $i$'s offspring takes over the site
$k$ is proportional to the player's fitness $f_i$ but decreases as 
$r_{ik}^{-\beta}$ where $r_{ik}$ is the distance between them.
In other words, $p_{ik}$ is given by
\beqa
 p_{ik} \a= \frac{1}{Z_k} \frac{e^{\omega \Pi_i}}{r_{ik}^\beta},
 \label{eq.pik}
\eeqa
where $Z_k \equiv \sum_{i\not=k}  e^{\omega \Pi_i} \times r_{ik}^{-\beta}$ is a
normalization factor. Note that the player at the death site $k$ cannot leave an
offspring. In particular, $s_k$ will become $C$ with probability
\beqa
\pkcz
\a= \frac{1}{Z_k}\,
\sum_{i\not=k} \frac{e^{\omega\Pi_i}}{r_{ik}^\beta}\, \delta_{s_i,C},
\eeqa
where $\delta_{x,y}$ is the Kronecker delta.
To remove trivial absorbing states where every player takes the same
action, we introduce a small mutation probability $\mu$. 
When player $i$ reproduces its offspring at the site $k$,
$s_k$ becomes $s_i$ with probability $1-\mu$ and by the opposite of $s_i$
with probability $\mu$. If we combine the DB process and mutation
together, the actual probability that $s_k$ becomes $C$ is given 
by
\beqa
\pkc
\a=  \pa{1-\mu} \pkcz + \mu \pa{1-\pkcz}.
\eeqa
It is straightforward to see that the complementary event $s_k \to D$
occurs with probability $1-\pkc$. 

We demonstrate the reproduction process in Fig.~\ref{f.model}(b) for
$\beta=2$ and $n=3$. 
When a player, say, $i=0$, is chosen for death, the
other players $k\not=0$ compete to produce their offspring at this
empty site. Player $k$'s winning probability is given by
$\frac{1}{|k|^2\,Z_0}\, e^{\omega \Pi_k}$ with
$\Pi_k$ given in the caption of (a). Here
$Z_0 = e^{\omega \Pi_{1}}+e^{\omega \Pi_{-1}}
+\frac{1}{4} \pas{e^{\omega \Pi_{2}}+e^{\omega \Pi_{-2}}}
+\frac{1}{9} \pas{e^{\omega \Pi_{3}}+e^{\omega \Pi_{-3}}}$
so that the normalization factor $1/Z_0$ makes
the total sum of probabilities be 1. The winner's offspring inherits
the winner's strategy with probability $(1-\mu)$ or the opposite one
with $\mu \ll 1$. 

\section{Analysis}
\label{sec:analysis}

We begin by considering how the threshold cost $\cth (\alpha, \beta)$
behaves in special cases. The first one is the case with the
nearest-neighbor competition, represented by $\beta \to \infty$.
The second one is when
$\alpha = \beta$, and $\cth (\alpha, \alpha)$ in the weak-selection limit
is readily obtained by applying the result of
Ref.~\onlinecite{allen2017evolutionary}.

\subsection{Nearest-neighbor competition ($\beta \to \infty$)}

\begin{figure}[t] 
\includegraphics[scale=0.7]{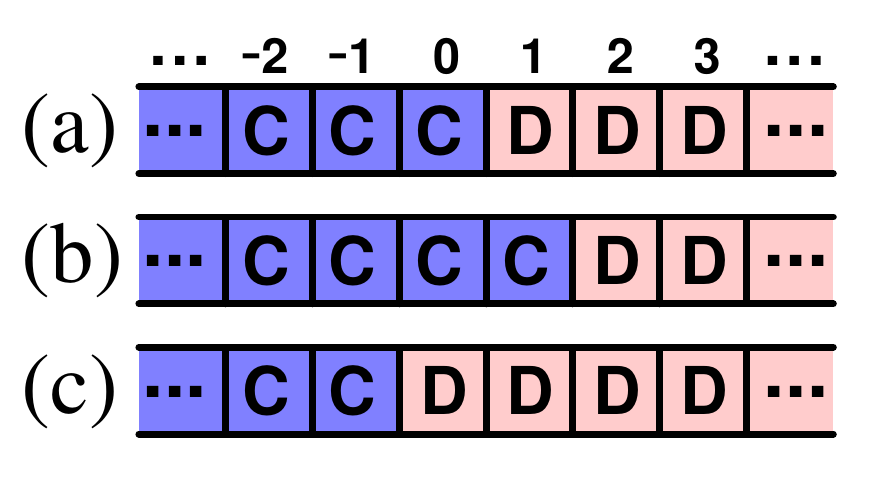}
\caption{(Color online)
  Domain-wall dynamics in case of the nearest-neighbor competition ($\beta \to
  \infty$).
  (a) Initial configuration in which two domains meet at a wall
  between the sites $k=0$ and $1$. The competition for reproduction
  occurs between the players at these sites.
  (b) If $k=1$ is made empty and occupied by an offspring of the player at
  $k=0$, the domain wall moves right.
  (c) On the contrary, the domain wall moves left if
  $k=0$ is made empty and taken over by an offspring of the player at $k=1$.
}
\label{f.bdmove}
\end{figure} 

If $\beta \to \infty$, only the nearest neighbors of the death site will compete
for reproduction. Let us define a domain as a set of consecutive sites with the
same action, bounded by sites with the opposite action. When a death site $k$ is
picked up inside a domain, therefore, the action of its neighbors must be the
same as $s_k$. If $\mu$ is negligibly small, we can say that the configuration
$\left\{ s_i \right\}$ will not change. It implies that the system shows
dynamics only at domain boundaries. Let us assume that the system is divided
into two equally large domains so that
\begin{equation}
s_i =
\left\{
\begin{array}{lcl}
C & \text{for} & i=0, -1, -2, \ldots, -n\\
D & \text{for} & i=1, 2, \ldots, n,
\end{array}
\right.
\end{equation}
as shown in Fig.~\ref{f.bdmove}(a). Assume that $k=1$ is chosen as for death.
Only its neighboring players $k=0$ and $k=2$ compete to take over the empty
site, and the domain wall will move right if the offspring of player $i=0$
wins. The payoffs $\Pi_0$ and $\Pi_2$ are calculated as
\beqa
\Pi_0
\a= \frac{1}{2\zeta_{n}(\alpha)}
\sum_{j\not=0}  \frac{\pi_{s_0 s_j}}{r_{0j}^\alpha}\nnn
\a= \frac{1}{2\zeta_{n}(\alpha)}\sum_{r=1}^{n}
\pa{\frac{1-c}{r^\alpha}+\frac{-c}{r^\alpha}}
\b= \frac{1}{2} - c, \\
\Pi_2
\a= \frac{1}{2\zeta_{n}(\alpha)} \sum_{j\not=2}  \frac{\pi_{s_2
s_j}}{r_{2j}^\alpha}\nnn
\a= \frac{1}{2\zeta_{n}(\alpha)} \sum_{r=2}^{n} \frac{1}{r^\alpha}
\b= \frac{1}{2} - \frac{1}{2\zeta_{n}(\alpha)},
\eeqa
and the probability that the wall moves right is given by
\beqa
P^0_{s_1 \to C}
\a= \frac{e^{\omega\Pi_0}}{e^{\omega\Pi_0}+e^{\omega\Pi_2}}.
\eeqa
Similarly, we can calculate the probability that the wall moves left,
$P^0_{s_0 \to D}$, from $\Pi_{-1}$ and $\Pi_{1}$.
The population will become cooperative when
$P^0_{s_1 \to C} > P^0_{s_0 \to D}$.
If $N \gg 1$, this inequality reduces to
\beqa
c < \frac{1}{2\zeta_{n}(\alpha)}
\label{e.cth}
\eeqa
and we have
$\cth (\alpha, \infty)=1/2\zeta_n(\alpha)$.

\subsection{Analytic solution of $\alpha = \beta$}

If the game interaction and competition share the same population
structure, strategies selected by the network reciprocity can be
obtained on any population structure in the weak-selection limit of
$\omega\to 0$ using
coalescent theory~\cite{allen2017evolutionary}.
When we denote the expected coalescence
time~\cite{wakeley2009coalescent}
from the two ends of an $n$-step random walk
by $\tau_n$, cooperation is favored when $c$ is smaller than
\begin{equation}
\cth = \frac{\tau_3-\tau_1}{\tau_2}
\label{eq.allen}
\end{equation}
in the weak-selection limit.
In a weighted undirected graph, if site $i$ is connected to $j$ by an
edge, the edge has a weight $w_{ij} = w_{ji}$. The weighted degree of
site $i$ is the sum of its edge weights:
$w_i \equiv \sum_j w_{ij}$. The probability for a random 
walker to transit from $i$ to $j$ is denoted by $p_{ij} \equiv w_{ij} / w_i$.
A useful quantity is the Simpson degree
$\kappa_i = \left( \sum_j p_{ij}^2 \right)^{-1}$,
a generalization of the topological degree.
For a regular graph, the site index $i$ is unnecessary since
every vertex has the same Simpson degree $k_i=k$. 
Equation~\eqref{eq.allen} is then rewritten as
\begin{equation}
\cth = \frac{N/\kappa - 2}{N - 2}.
\end{equation}
In our case, noting that $w_{ij} \propto r_{ij}^{-\alpha}$ for $i\neq j$ and
$w_{ii} = 0$, we find the threshold as a combination of the
zeta functions~\cite{hardy1979introduction},
\begin{equation}
\cth (\alpha, \alpha) = \frac{\zeta_{n}(2\alpha)}{2\pas{\zeta_{n}
\pa{\alpha}}^2},
\end{equation}
where we have taken into account $N \gg 1$ in the sense of the
$\omega N$-limit~\cite{jeong2014optional}.

\subsection{Conjecture for $\cth(\alpha, \beta)$}
To recap, we have two predictions:
First, if $\beta \to \infty$ and $N \gg 1$,
the threshold cost is predicted to be
\begin{equation}
\cth (\alpha, \infty) =\frac{1}{2\zeta_{n}(\alpha)},
\label{e.clue1}
\end{equation}
so that $\zeta_{n}(\alpha)$ can be interpreted as the effective
number of neighbors to play the game with on one side of the focal player.
Second, when $\alpha = \beta$,
the threshold cost $\cth$ has been derived as
\begin{equation}
\cth (\alpha, \alpha) =
\frac{\zeta_{n}(\alpha + \alpha)}{2\zeta_{n}(\alpha)\zeta_{n}(\alpha)}
\label{e.clue2}
\end{equation}
for $\omega \to 0$. To incorporate Eqs.~\eqref{e.clue1} and \eqref{e.clue2},
we conjecture a connection formula $\cth (\alpha, \beta)$ as follows:
\beqa
\cth (\alpha, \beta) =
\frac{\zeta_{n}(\alpha + \beta)}{2 \zeta_{n}(\alpha)\zeta_{n}(\beta)}.
\label{e.conj}
\eeqa
In particular, for $\alpha \to \infty$, the conjecture predicts
\begin{equation}
\cth (\infty, \beta) = \frac{1}{2\zeta_{n}(\beta)}.
\label{e.conjb}
\end{equation}

\section{Numerical calculation}
\label{sec:numeric}

For Monte Carlo simulation, we construct a cycle of size $N$ with a player at
each site. Initial $s_i$'s are randomly chosen between $C$ and $D$ equally
probably. Then, according to the DB process, we randomly select a player (say,
$k$) for death and choose another player $i$ with probability $p_{ik}$ in
Eq.~\eqref{eq.pik}.
One Monte Carlo step consists of $N$ trials to update the configuration
$\{s_i \}$. In addition, we introduce mutation with small probability
$\mu \ll 1$ so that the system has a unique stationary
distribution over configurations $\left\{ s_i \right\}$. The quantity of
interest is cooperator abundance $\xc$, the average frequency of cooperators in
the steady state. For each data point in a sample, we run the Monte Carlo
simulation for $6\times 10^3$ Monte Carlo steps, and the time average of $\xc$
is taken for the last $10^2$ Monte Carlo steps, at which the simulation has
already reached a steady state. The total number of samples amounts to $4\times
10^3$.

\subsection{Weak selection ($w=0.02$)}

\begin{figure}
\includegraphics[width=0.49\columnwidth]{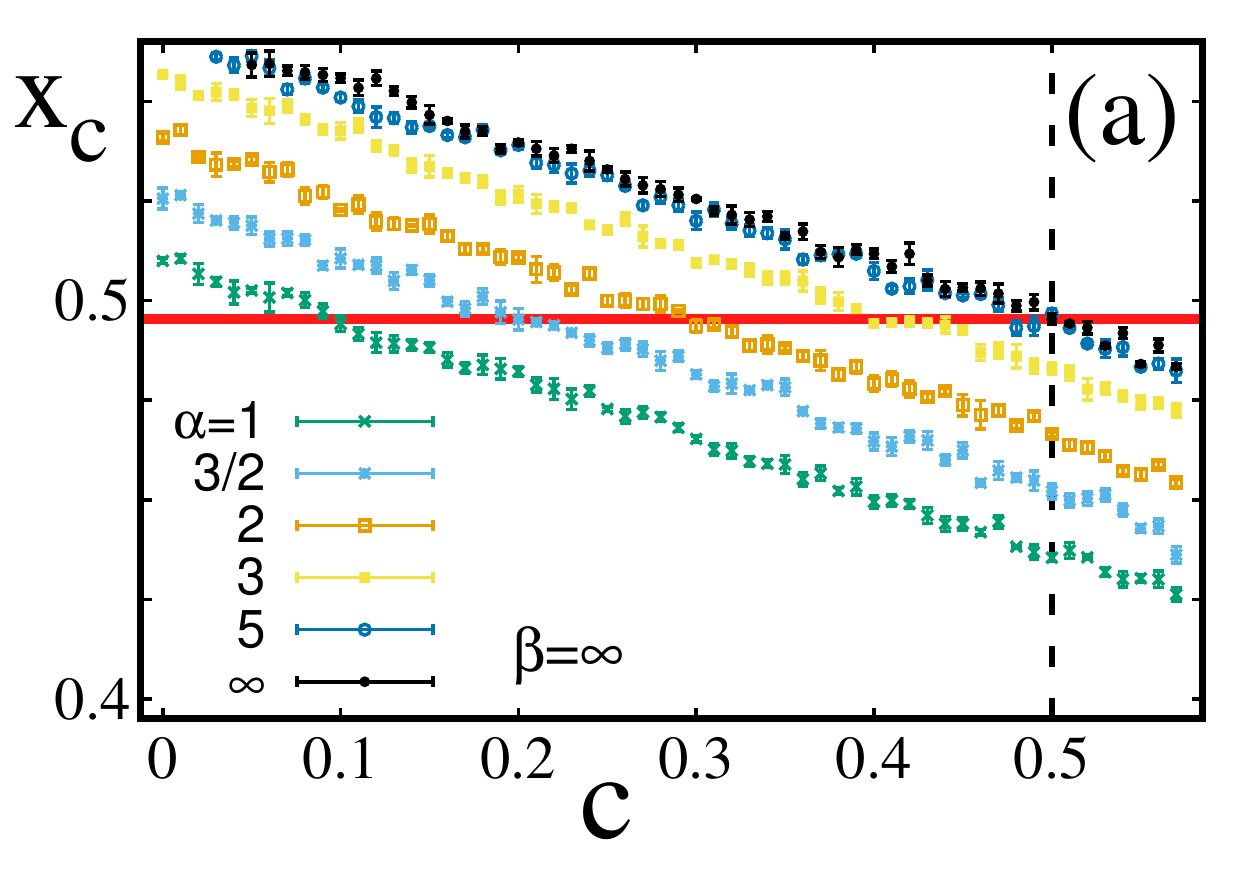}
\includegraphics[width=0.49\columnwidth]{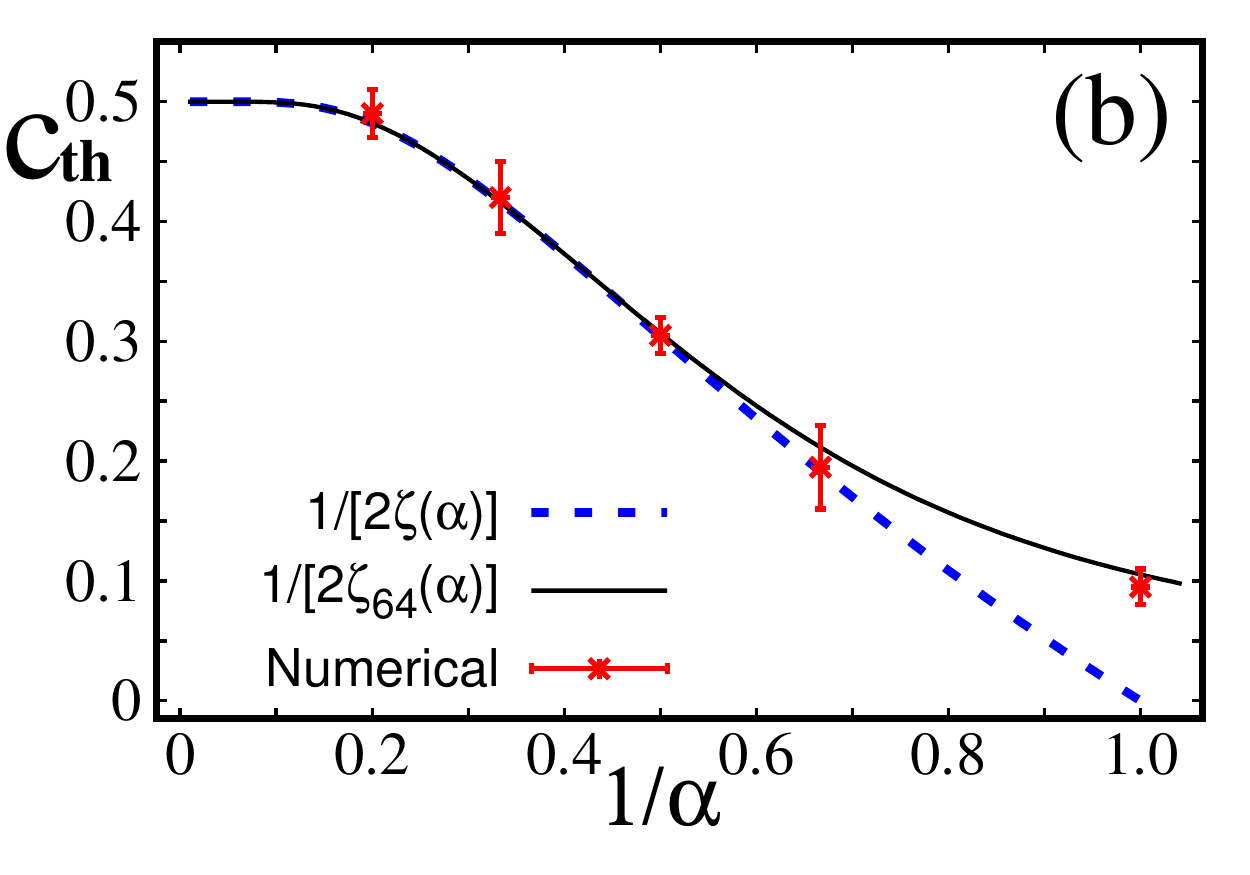}
\caption{(Color online)
Threshold estimation in case of nearest-neighbor competition
 ($\beta\to \infty$). The parameters for Monte Carlo simulation are
the total number of sites $N = 128$, selection strength
$\omega = 0.02$, and mutation rate $\mu =10^{-3}$.
(a) Cooperator abundance $\xc$ as a function of $c$ for each
different $\alpha$. The vertical dotted line means $c = 1/2$, and the
horizontal shaded area represents $0.495 \pm 0.001$, the error-bar range of
$\xc (c=1/2)$ for $\alpha \to \infty$.
(b) The error-bars show
$[c_{\min}, c_{\max}]$ for different values of $\alpha$. The solid
line depicts our analytic threshold in Eq.~\eqref{e.clue1} for
$N=128$, and the dotted line shows the limit of $N \to \infty$.}
\label{f.b200}
\end{figure}

\begin{figure}
\includegraphics[width=0.49\columnwidth]{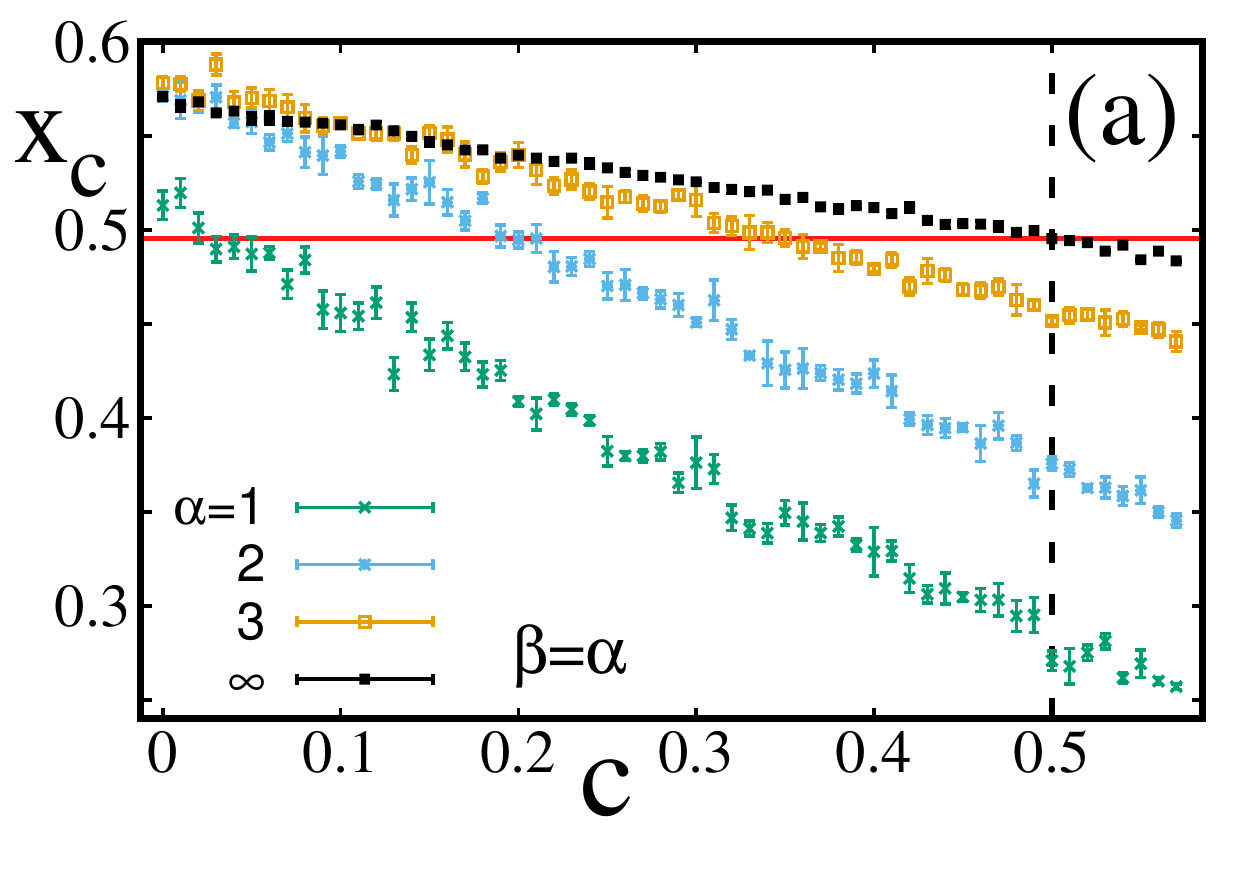}
\includegraphics[width=0.49\columnwidth]{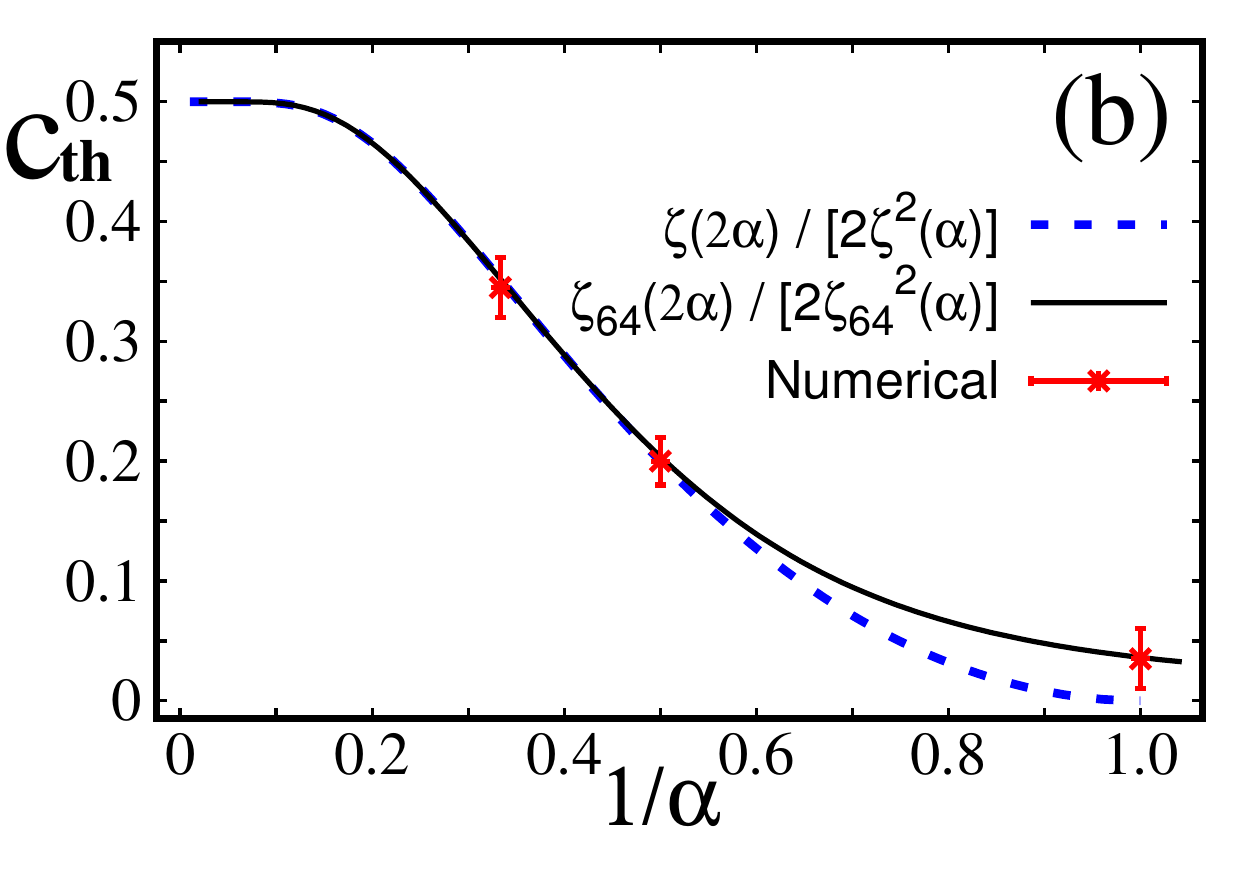}
\caption{(Color online)
Threshold estimation for $\alpha = \beta$. The simulation parameters
are the same as in Fig.~\ref{f.b200}. (a) Cooperator abundance $\xc$ as a
function of $c$ for each different $\alpha$. The vertical dotted line means $c =
1/2$, and the horizontal shaded area represents $0.495 \pm 0.001$, the error-bar
range of $\xc (c=1/2)$ for $\alpha \to \infty$. (b) The error-bars show
$[c_{\min}, c_{\max}]$ for different values of $\alpha$. The solid line depicts
our analytic threshold in Eq.~\eqref{e.clue2} for $N=128$, and the dotted line
shows the limit of $N \to \infty$.}
\label{f.ab}
\end{figure}

Noting that most of the analytic results have been obtained in the
weak-selection limit, we have to check this limit first and proceed to stronger
selection strengths thereafter.
Let us begin by setting $\beta \to \infty$. If $\alpha$ also goes to infinity,
the threshold at which $\xc(\cth) = 1/2$ is analytically given as
$\cth (\infty, \infty) = 1/2$ in the $\omega N$-limit~\cite{jeong2014optional}.
However, this limit cannot be taken exactly in the Monte Carlo simulation,
and $\mu$ cannot be infinitesimally small either. For $c=1/2$,
we actually observe $\xc = 0.495 \pm 0.001$. We regard this value as a
reference point to estimate $\cth(\alpha,\infty)$, assuming that it takes into
account all the corrections due to the difference from the $\omega N$-limit as
well as the finiteness of $\mu$.
The procedure to estimate $\cth (\alpha,\infty)$ now goes as follows: For given
$\alpha$, we calculate $\xc$ as a function of $c$, together with the standard
error from statistically independent samples. As a lower bound of $\cth (\alpha,
\infty)$, we pick up $c_{\min}$, defined as
$c$ below which all the error bars of $\xc (c)$ lie above $0.495 \pm
0.001$. Likewise, the upper bound of $\cth (\alpha, \infty)$ is given by
$c_{\max}$, defined as $c$ above which all the error bars of $\xc (c)$ lie
below $0.495 \pm 0.001$ [Fig.~\ref{f.b200}(a)]. The numerical estimate of $\cth
(\alpha, \infty)$ is thus represented by an interval $[c_{\min}, c_{\max}]$ as
shown in Fig.~\ref{f.b200}(b). The agreement with Eq.~\eqref{e.clue1} is
compelling.
The above procedure can be applied to the case of $\alpha = \beta$ as well
[Fig.~\ref{f.ab}(a)]. Here, the estimation is more difficult
because the slope of $\xc$ becomes steeper as $\alpha$ decreases. Nevertheless,
Fig.~\ref{f.ab}(b) shows that our estimates are
consistent with the analytic solution [Eq.~\eqref{e.clue2}].

\begin{figure}
\includegraphics[width=0.49\columnwidth]{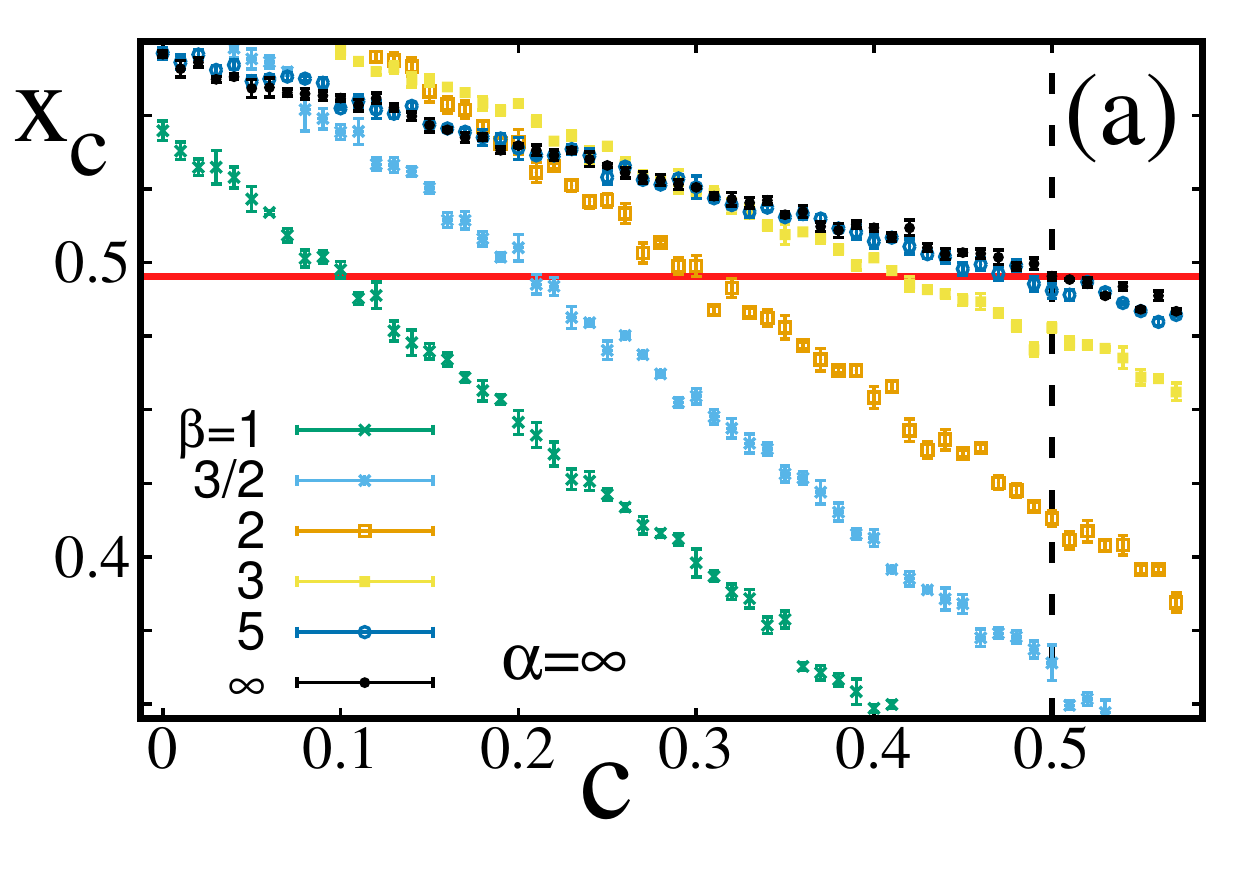}
\includegraphics[width=0.49\columnwidth]{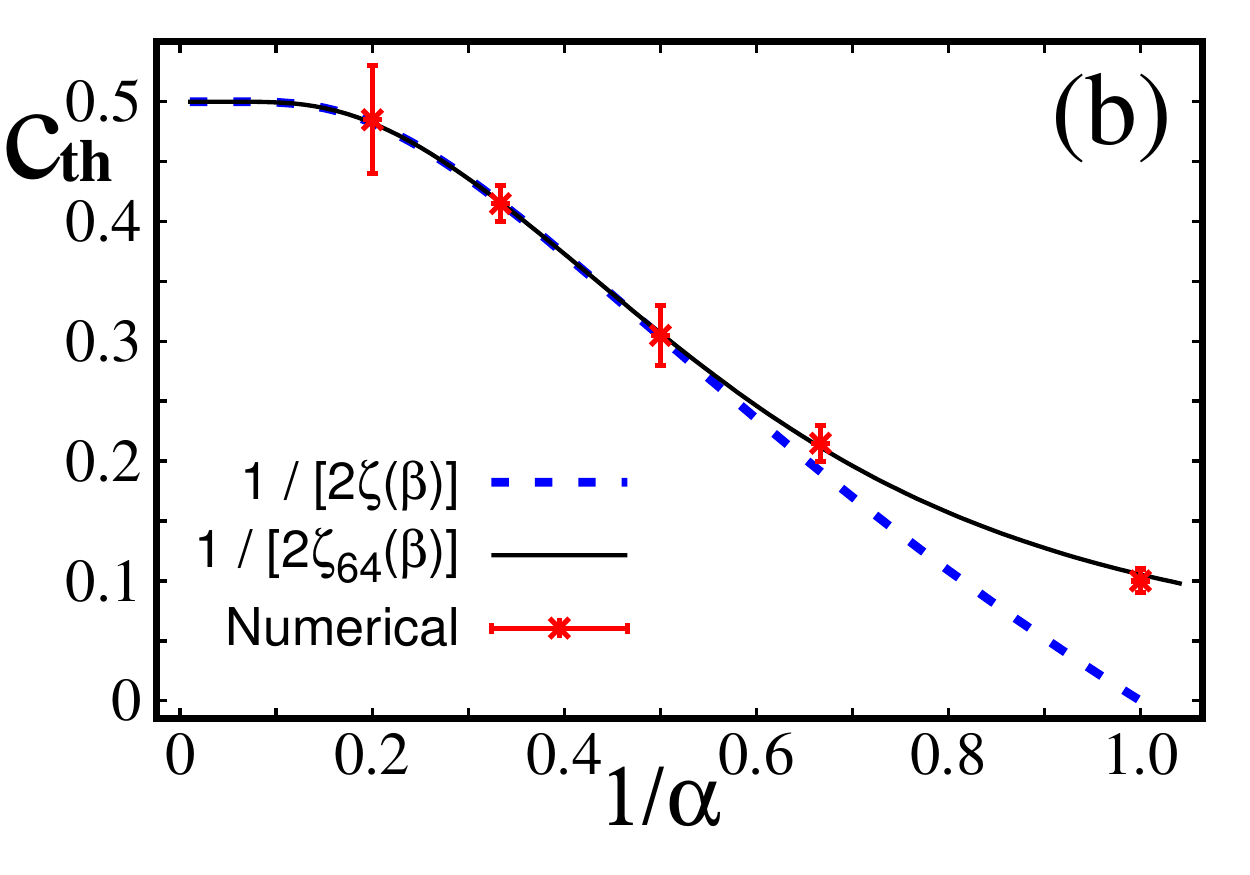}
\caption{(Color online)
Threshold estimation for $\alpha = \infty$. The simulation parameters
are the same as in Fig.~\ref{f.b200}. (a) Cooperator abundance $\xc$ as a
function of $c$ for each different $\beta$. The vertical dotted line means $c =
1/2$, and the horizontal shaded area represents $0.495 \pm 0.001$, the error-bar
range of $\xc (c=1/2)$ for $\beta \to \infty$. (b) The error-bars show
$[c_{\min}, c_{\max}]$ for different values of $\beta$. The solid line depicts
our conjectured threshold in Eq.~\eqref{e.conjb} for $N=128$, and the dotted
line shows the limit of $N \to \infty$.}
\label{f.a200}
\end{figure}

\begin{figure}
\includegraphics[width=0.49\columnwidth]{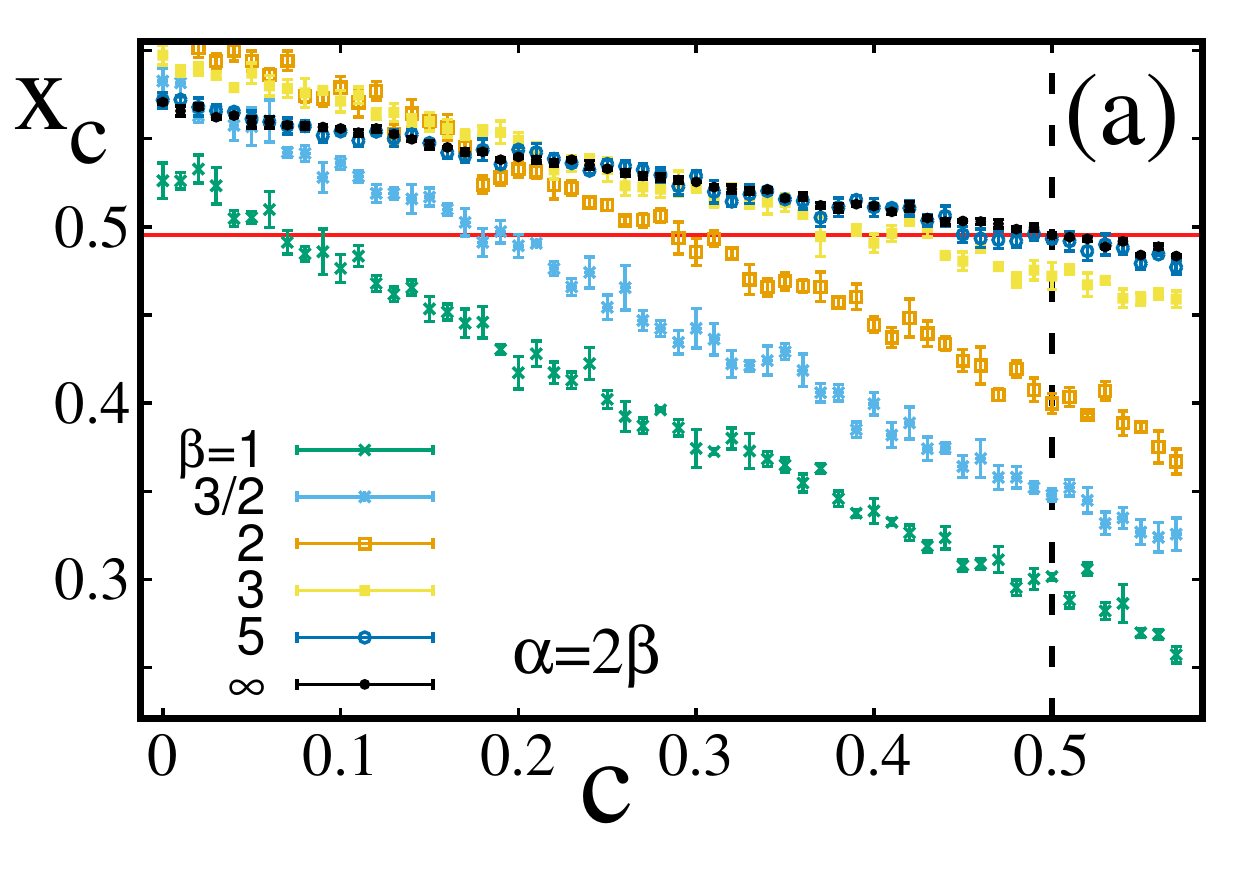}
\includegraphics[width=0.49\columnwidth]{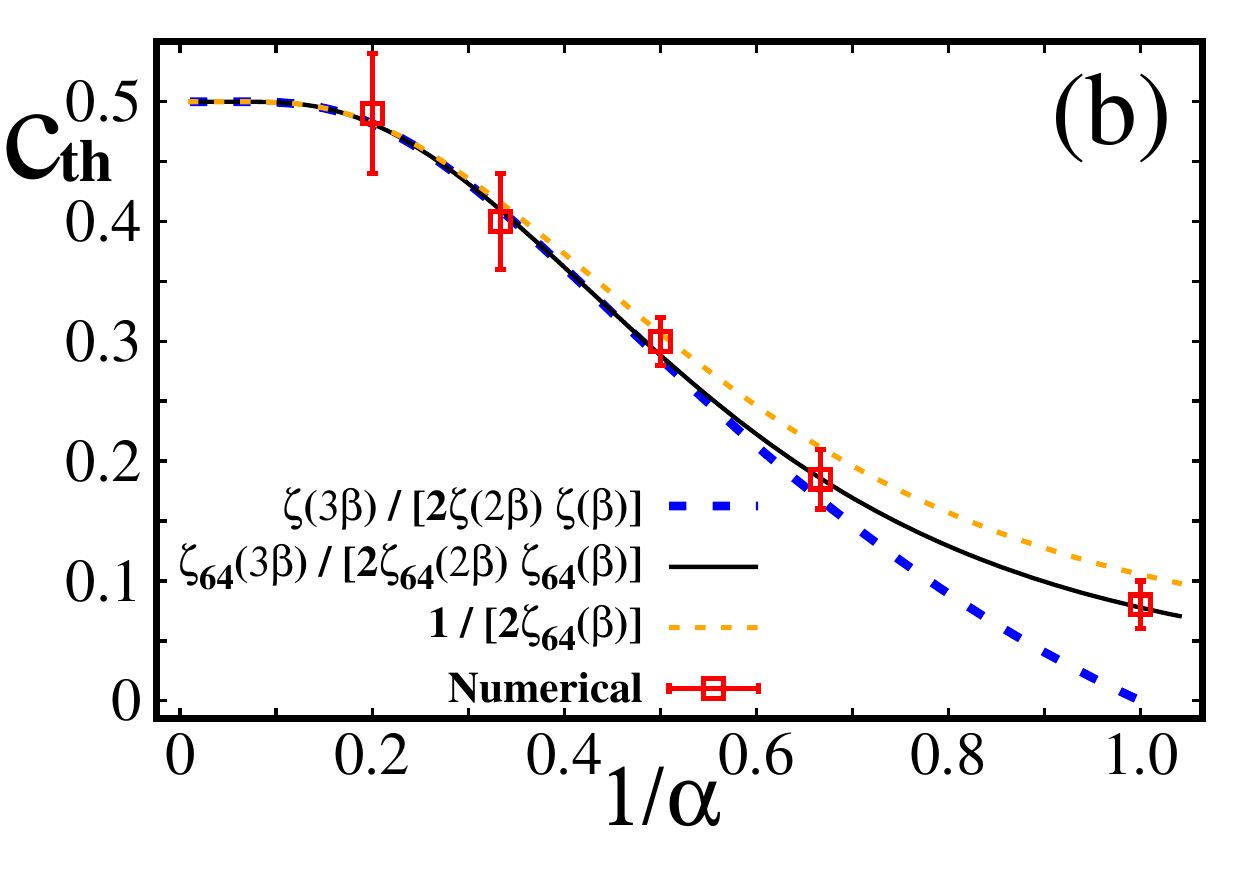}
\caption{(Color online)
Threshold estimation for $\alpha = 2\beta$. The simulation parameters
are the same as in Fig.~\ref{f.b200}. (a) Cooperator abundance $\xc$ as a
function of $c$ for each different $\beta$. The vertical dotted line means $c =
1/2$, and the horizontal shaded area represents $0.495 \pm 0.001$, the error-bar
range of $\xc (c=1/2)$ for $\beta \to \infty$. (b) The error-bars show
$[c_{\min}, c_{\max}]$ for different values of $\beta$. The solid line depicts
our conjectured threshold in Eq.~\eqref{e.conj} for $N=128$, and the dotted
line shows the limit of $N \to \infty$. For comparison, we have also plotted
$1/[2\zeta_{64}(\beta)]$ as in Fig.~\ref{f.a200}(b), which deviates from the
numerical data.}
\label{f.a2b1}
\end{figure}

Let us move on to check whether the conjecture of Eq.~\eqref{e.conj} correctly
predicts the threshold behavior even when no analytic proof is available.
One specific example that we can test is an extreme case of $\alpha \to \infty$,
where the conjecture reduces to Eq.~\eqref{e.conjb}.
Comparing Fig.~\ref{f.a200}(a) with Fig.~\ref{f.b200}(a),
we see that $\alpha$ and $\beta$ do not play the same role in determining the
cooperator abundance. Nevertheless, Fig.~\ref{f.a200}(b) substantiates the
conjectured symmetry between $\alpha$ and $\beta$ in the threshold behavior.
To make both $\alpha$ and $\beta$ finite yet unequal, we consider another case
of $\alpha = 2\beta$ [Fig.~\ref{f.a2b1}(a)]. Again, our numerical results
support the conjecture in Eq.~\eqref{e.conj}. The overall behavior might look
similar to the previous case of $\alpha=\infty$, but a closer look shows that
our numerical estimates are sharp enough to distinguish the difference between
them [Fig.~\ref{f.a2b1}(b)].

\subsection{Stronger selection ($w=1$ and $w=5$)}

\begin{figure}
\includegraphics[width=0.49\columnwidth]{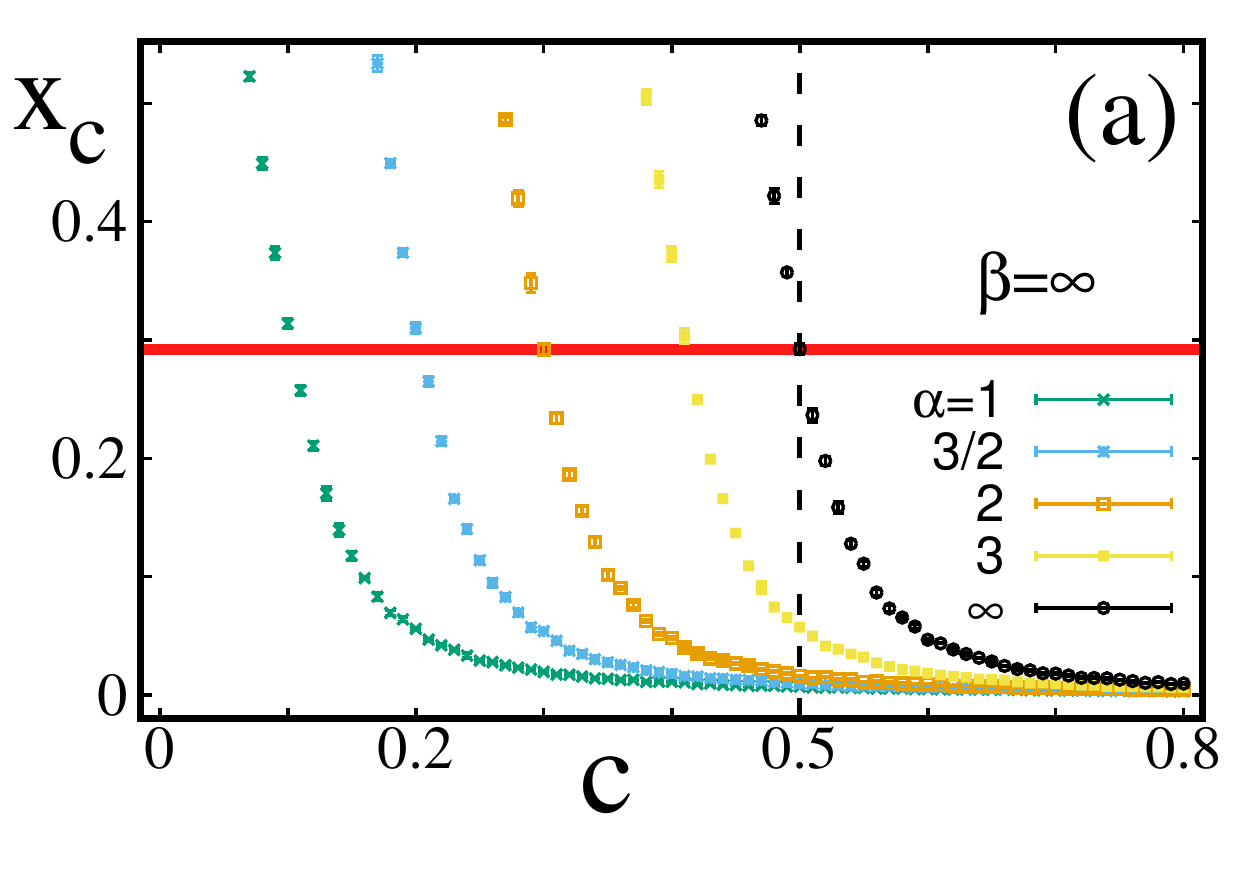}
\includegraphics[width=0.49\columnwidth]{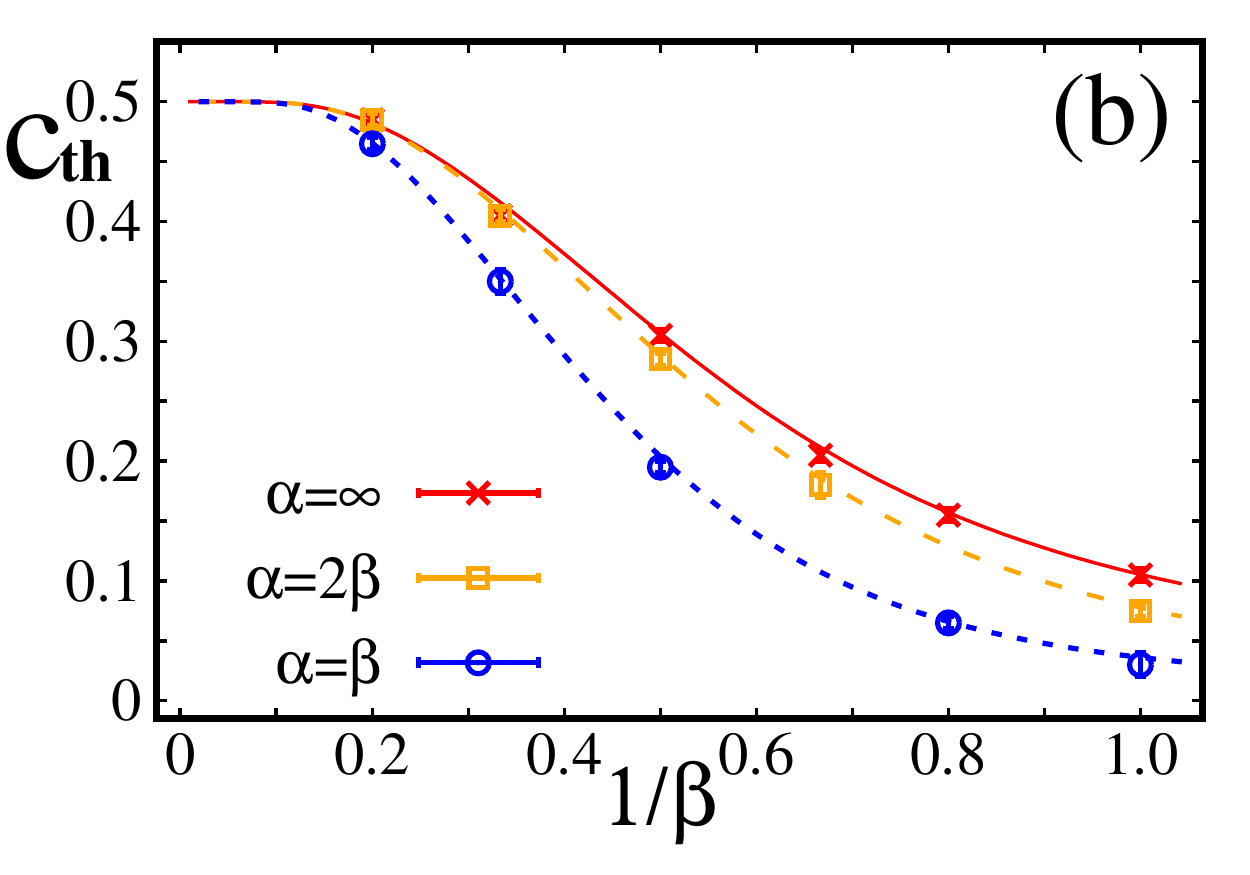}
\caption{(Color online)
Threshold estimation for $w=1$. The other simulation parameters
are the same as in Fig.~\ref{f.b200}. (a) Cooperator abundance $\xc$ as a
function of $c$ for each different $\alpha$, while $\beta \to \infty$.
The vertical dotted line means $c =
1/2$, and the horizontal shaded area represents $0.2923 \pm 0.0045$, the
error-bar range of $\xc (c=1/2)$ for $\alpha \to \infty$. (b) The error-bars
show $[c_{\min}, c_{\max}]$ for different values of $\beta$. The lines
depict our conjectured thresholds in Eq.~\eqref{e.conj}.}
\label{f.w1}
\end{figure}

\begin{figure}
\includegraphics[width=0.49\columnwidth]{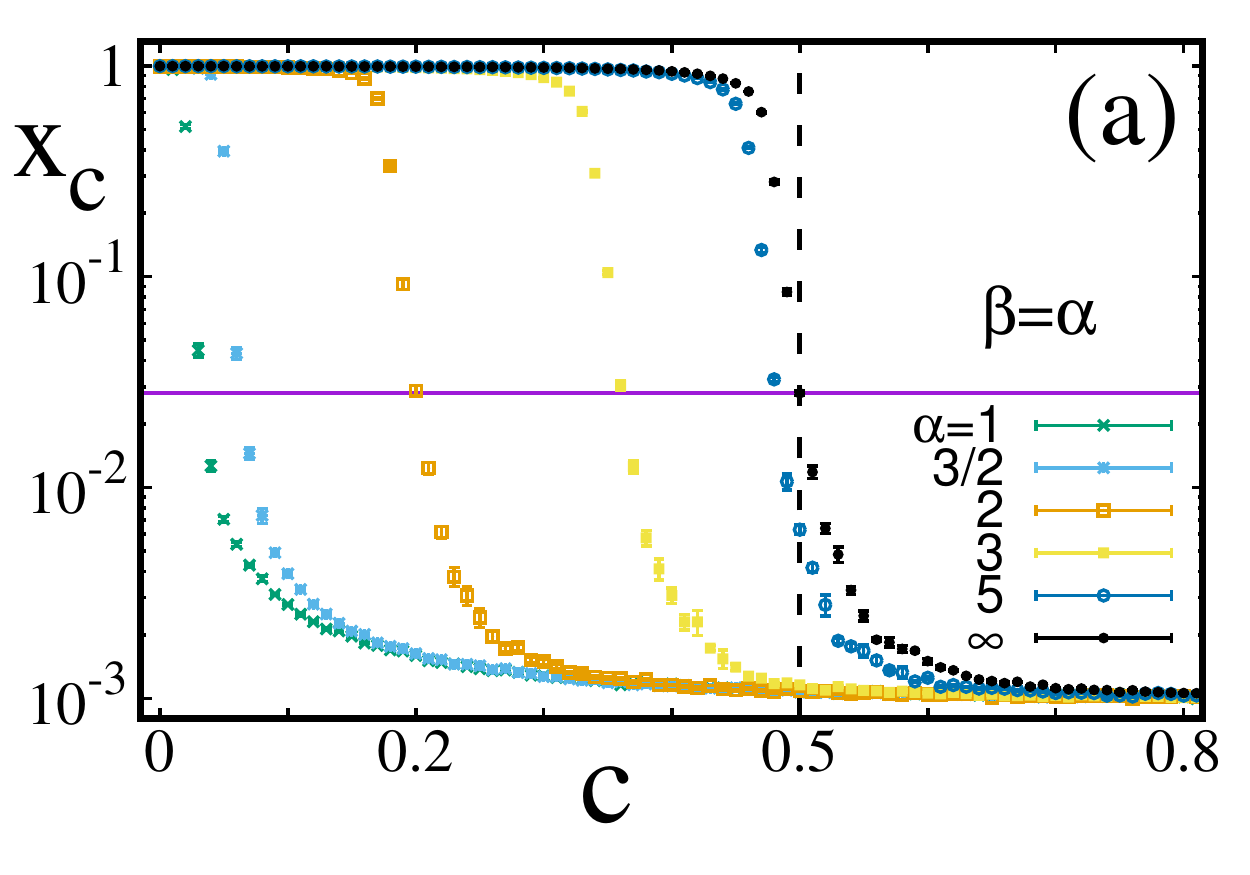}
\includegraphics[width=0.49\columnwidth]{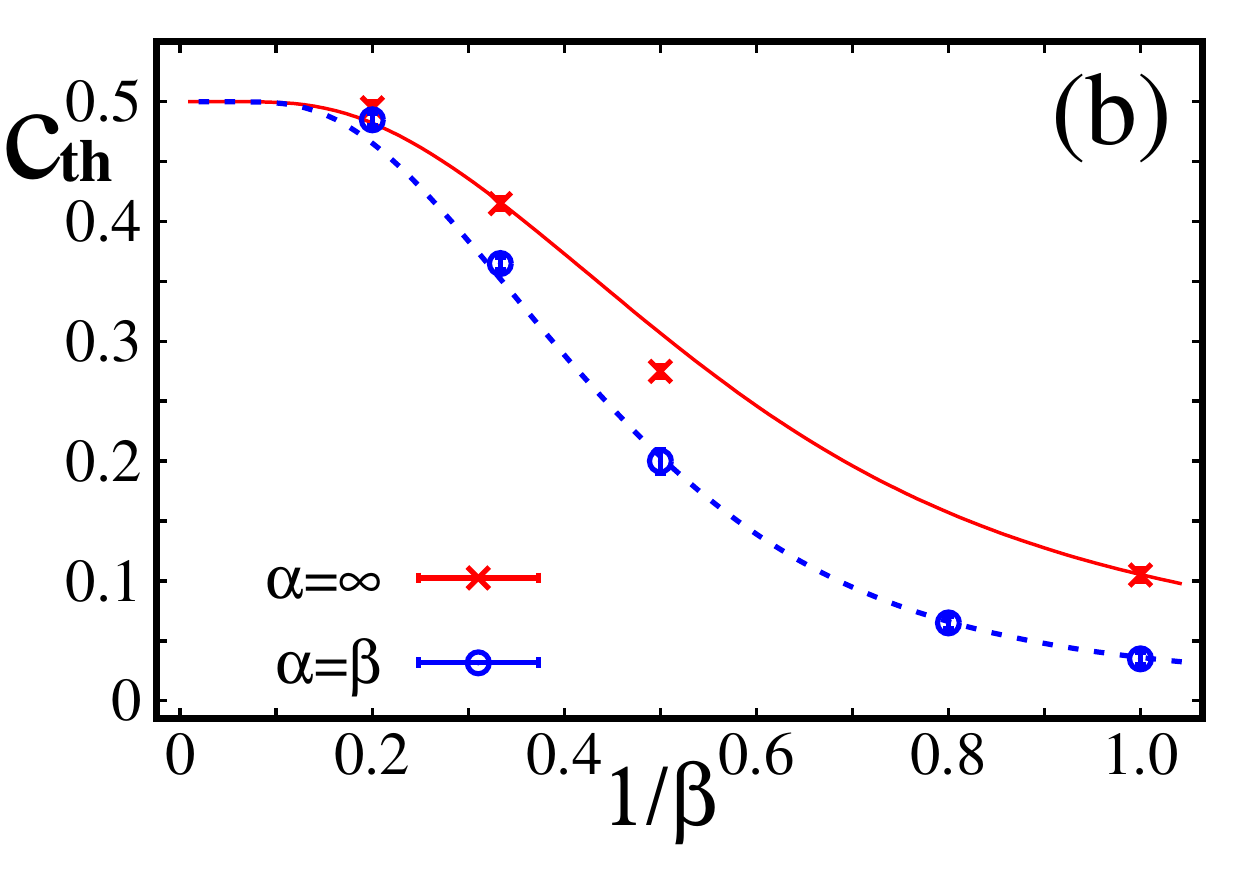}
\caption{(Color online)
Threshold estimation for $w=5$. The other simulation parameters
are the same as in Fig.~\ref{f.b200}. (a) Cooperator abundance $\xc$ as a
function of $c$ for each different $\alpha = \beta$. The vertical dotted line
means $c = 1/2$, and the horizontal shaded area represents $0.0280 \pm 0.0007$,
the error-bar range of $\xc (c=1/2)$ for $\alpha \to \infty$. Note the
logarithmic scale on the vertical axis. (b) The error-bars
show $[c_{\min}, c_{\max}]$ for different values of $\beta$. The lines
depict our conjectured thresholds in Eq.~\eqref{e.conj}.}
\label{f.w5}
\end{figure}

The conjectured formula [Eq.~\eqref{e.conj}] works even beyond the
weak-selection limit. Our numerical estimation procedure hinges on the fact that
we can still take $c=1/2$ as the threshold in a double limit of
$\alpha \to \infty$ and $\beta \to \infty$ for arbitrary selection
strength~\cite{jeong2014optional}. In this
limit, our numerical calculation gives $x_c = 0.2923 \pm 0.0045$ at $c=1/2$,
provided that we have set $w=1$ [Fig.~\ref{f.w1}(a)]. As
we have done above, we will take this value as a reference point to locate the
threshold in other combinations of $\alpha$ and $\beta$ for which one can
suitably define an extrapolation to $\alpha \to \infty$ and $\beta \to \infty$:
Let us take a look at the results shown in Fig.~\ref{f.w1}(b). Here, we
demonstrate three cases, $\alpha = \infty$, $\alpha=2\beta$, and $\alpha=\beta$.
Note that the limit can be approached by taking $\beta \to \infty$ for all these
cases, so that the abundance levels may be compared with the reference value in
the double limit. Although none of the cases have been solved analytically, the
threshold behavior is indeed captured by Eq.~\eqref{e.conj} with high
precision.

As $w$ increases, the quality of fit deteriorates, however.
For $w=5$, for example, the reference value is lowered to $x_c = 0.0280 \pm
0.0007$ [Fig.~\ref{f.w5}(a)]. It is a tiny value, and the slope of $\xc(c)$ is
steep, which makes the estimate more challenging than the previous
weaker-selection cases. It is not surprising that deviations from the conjecture
become visible [Fig.~\ref{f.w5}(b)], but the overall agreement is still
remarkable.

\section{Concluding remarks}
\label{sec:conclude}

To summarize, we have studied the PD game on a cycle, where the
probabilities for two players at distance $r$ to play the game and to
compete for reproduction decay algebraically as $r^{-\alpha}$ and
$r^{-\beta}$, respectively. When the benefit of cooperation is fixed
as unity, there exists a threshold cost $\cth (\alpha, \beta)$ below
which the population becomes cooperative. Based on analytically
tractable cases [Eqs.~\eqref{e.clue1} and \eqref{e.clue2}], we have
conjectured its functional form as given in Eq.~\eqref{e.conj}.
The apparent symmetry between $\alpha$ and $\beta$ in Eq.~\eqref{e.conj} is
nontrivial as one sees that Fig.~\ref{f.b200}(a) clearly
differs from~\ref{f.a200}(a).
Although we are unable to prove our conjecture at the moment,
the striking agreement with numerical results suggests that the exact
result of $c(\alpha,\alpha)$ can be generalized in a simple form, at least
in this particular long-ranged interaction model on a cycle.

The threshold costs in our model do not seem to depend on the
selection strength $\omega$ strongly. For the nearest neighbor competitions
($\beta\to\infty$), this independence can be understood easily since the
threshold cost is obtained from the domain boundary dynamics
as discussed in~III A. For finite $\beta$, our conjecture relies on the
analytic solution of weak selection limit. However, the threshold
costs are almost independent of selection strength up to
$\omega \approx 5$.
Further studies on the robustness of the weak selection
limit as well as the threshold costs for general 
population structures are needed.

\acknowledgments
S.K.B. was supported by Basic Science Research Program through the
National Research Foundation of Korea (NRF) funded by the Ministry of Science,
ICT and Future Planning (NRF-2017R1A1A1A05001482). H.C.J. was
supported by Basic Science Research Program through the National
Research Foundation of Korea (NRF) funded by the Ministry of Education
(NRF-2018R1D1A1A02086101).

%


\end{document}